# PBG properties of three-component 2D photonic crystals.

Alexander Glushko* and Lyudmila Karachevtseva

*Institute of Semiconductor Physics of NAS of Ukraine; 41 Nauki Prsp., Kiev-03028, Ukraine;*

*Tel.:(+38-044)525-98-15, Fax: (+38-044)525-83-42*

* Corresponding author. E-mail address: a_glushko@ukr.net

**Abstract**.

In this paper we analyze theoretically how the introduction of the third component into the two-dimensional photonic crystal influences the photonic band structure and the density-of-states of the system. We consider the periodic array of cylindrical air rods in a dielectric, and the third medium is introduced as a ring-shaped intermediate layer of thickness d and dielectric constant $\varepsilon_i$ between the air pores and the dielectric background. Using the plane wave method, we have obtained the band structures for the 2D triangular lattice photonic crystals. The dependencies of TE and TM band gaps' widths and gaps' edges position on the interlayer dielectric constant and interlayer thickness were analyzed. In the framework of this approach, we have estimated the influence of the surface oxide layer on the band structure of macroporous silicon. We observed the shift of the gaps' edges to the higher or lower frequencies, depending on the interlayer thickness and dielectric constant. We have shown that the existence of a native oxide surface layer should be taken into consideration to understand the optical properties of 2D photonic crystals, particularly in macroporous silicon structures.

Keywords: surface layer; band gap transformation; tunable photonic crystal.

## 1. Introduction

Photonic band gap materials have excited a great interest during the last decade, due to their potential applications in all-optical logical elements, high-efficiency lasers fabrication, low-loss waveguides, and other fields. Photonic crystals of various configurations, symmetries and compositions were introduced since the beginning of the 1990s, when the first theoretical and experimental methods for photonic crystals fabrication and investigations were established [1-10]. One of the recent directions of the investigations is the fabrication of photonic crystals, which allow tuning of the photonic band gap edges by an external influence [11-17].

In this paper, we report on a theoretical study of the three-component 2D photonic crystals. It is considered the array of cylindrical rods with dielectric constant (DC) $\varepsilon_a$ embedded into the dielectric with DC $\varepsilon_b$. The third medium is introduced as a ring-shaped intermediate layer around the rod's surface with DC $\varepsilon_i$ (Fig.1).

Interest in the problem appears from the experimental results for macroporous silicon structures obtained by the method of electrochemical etching [4]. The near-surface region of the sample fabricated by this method has properties different from the bulk ones [8]. To our knowledge, the influence of these specific regions onto the photonic band structure was not investigated earlier. Silicon is a very efficient material for fabrication of photonic crystals, due to its high dielectric constant $\varepsilon=11.7$ in the transparency band lying below the frequency of electronic band gap and variety of possible 2D and 3D configurations [18]. One more attractive property of silicon is the third-order nonlinearity which can be induced by free-carrier injection [14, 17, 19-21]. A nonlinear photonic crystal gives a possibility to control the light propagation by fitting the intensity of the pump beam. In this case, the frequency of the signal should lie in the vicinity of the band gap edge, since a nonlinear shift of the refractive index and, consequently, shift of the band gap position are very small (in [14], of the order of 30 nm). But if a device works not in a vacuum, the photonic crystal's "atoms" may slightly change their properties near the interfaces due to some chemical processes. For example, the inner pore's surface of a macroporous silicon-air structure inevitably contains a $SiO_2$ interlayer. On the other hand, intentional oxidization of the macroporous silicon is widely used, too [18]. Thus, the aim of our work is to estimate how covering the 2D photonic crystal's "atoms" by some dielectric lossless material will influence on the photonic band structure.

We use the plane wave method (PWM) to calculate the electromagnetic wave propagation through the 2D dielectric lossless photonic crystals. PWM was described in details in several papers [2, 3 ,5-7] and we use the notations introduced by Maradudin and co-authors [3, 7, 10]. The direction of propagation of the electromagnetic waves in all the cases is in the plane $x_1x_2$ (Fig. 1). We used the basis of Fourier expansion of 729 (inaccuracy is near 2%) vectors of the reciprocal lattice with the exception of the data in Table 1, where the basis was 1519 (inaccuracy is near 0.5%) vectors.

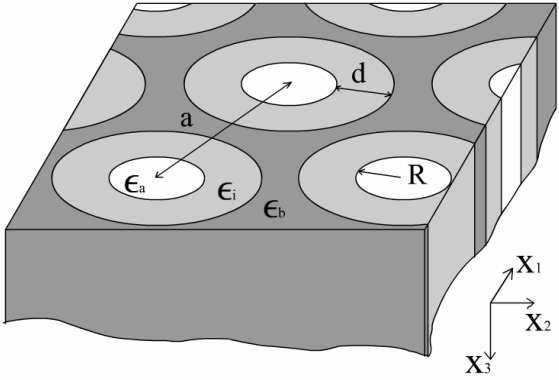

Fig. 1. Fragment of the system we consider. $\varepsilon_a$, $\varepsilon_b$, $\varepsilon_i$ are the dielectric constants of the rod, background, and interlayer, respectively. $R$ is the rods radius, $d$ is the interlayer thickness, and $a$ is the lattice constant.

The PWM gives only eigen electromagnetic states of a photonic crystal with periodic boundary conditions in $x_1x_2$ plane and infinite media in the $x_3$ direction. Thus, all the predictions we will make concern the structure with a sufficiently high aspect ratio of the pores. Nevertheless, we believe that at least a qualitative description of the influence of the third component in 2D photonic crystal on the photonic band structure will be in good agreement with the experiment.

## 2. Results and Discussion

Let us consider the 2D triangular lattice photonic crystal depicted in Fig.1. An additional layer on the pore's surface is characterized by two parameters: DC $\varepsilon_i$ and thickness $d$. We restrict ourselves to consideration of the air pores $\varepsilon_a=1$ in the dielectric with $\varepsilon_b=12$. Thus, the band structure of our three-component photonic crystal is determined by three parameters: $R$, $\varepsilon_i$, and $d$. Evidently, the insertion of such a layer does not change the symmetry of the system. All information about the dielectric constants and filling fractions of a photonic crystal is contained in Fourier coefficients of the reciprocal dielectric function, which, in our case, are given by

$$\hat{\kappa}(G_{II}) = \begin{cases} \dfrac{2\cdot\left(\dfrac{1}{\varepsilon_a}-\dfrac{1}{\varepsilon_b}\right)\cdot f_1 J_1(G_{II} R)}{G_{II} R} + \dfrac{2\cdot\left(\dfrac{1}{\varepsilon_i}-\dfrac{1}{\varepsilon_b}\right)\cdot f_3 J_1(G_{II}\cdot(R+d))}{G_{II}(R+d)}, & G_{II} \neq 0 \\ \left(\dfrac{f_1}{\varepsilon_a}+\dfrac{f_2}{\varepsilon_b}+\dfrac{f_i}{\varepsilon_i}\right), & G_{II} = 0 \end{cases} \quad (1)$$

where $f_1 = \frac{2\pi}{\sqrt{3}} \frac{R^2}{a^2}$, $f_i = \frac{2\pi}{\sqrt{3}} \frac{((R+d)^2 - R^2)}{a^2}$, $f_3 = \frac{2\pi}{\sqrt{3}} \frac{(R+d)^2}{a^2}$, $f_2 = 1 - f_3$, $J_1(x)$ is the Bessel function, and $a$ is the distance between the nearest-neighbors cylinders (lattice pitch). Note, in limiting cases, $\varepsilon_i=\varepsilon_a$, $\varepsilon_i=\varepsilon_b$, $d=0$, $R=0$, the expression (1) transforms into the expression for the usual 2D photonic crystal [3, 6, 7]:

$$\widehat{\kappa}(\vec{G}_{II}) = \begin{cases} f\frac{1}{\varepsilon_a} + (1-f)\frac{1}{\varepsilon_b}, & \vec{G}_{II} = 0 \\ \left(\frac{1}{\varepsilon_a} - \frac{1}{\varepsilon_b}\right) f \frac{2J_1(G_{II}R)}{G_{II}R}, & \vec{G}_{II} \neq 0 \end{cases} \quad (2)$$

where $f = \frac{2\pi}{\sqrt{3}} \frac{R^2}{a^2}$.

It is necessary to explore here the question concerning the accuracy of the three-component PWM. Intuitively, one can suppose that convergence of the three-component PWM will be poorer than for the two-component one. We have calculated the dependencies of TE and TM band gaps' edges positions on the number of basis reciprocal vectors up to $N=1519$. All curves show good convergence. The difference between the values calculated, for instance, for $N=1387$ and $N=1519$ is less than *0.09%*, thus further increasing of the basis seems senseless. We believe that the accuracy of the results described below is sufficient: *~2%* (*N=729*) for the plots in Figs. 2 and 3; *~0.5%* (*N=1519*) for the data in Table 1.

Figure 2 shows the gap maps for the lowest TE (dashed) and TM (solid) gaps in the triangular lattice photonic crystal with $R=0.42a$ depending on the thickness of interlayer with dielectric constant *a)* $\varepsilon_i=1$, *b)* $\varepsilon_i=2$, *c)* $\varepsilon_i=24$. In fact, case *a)* corresponds to the pore's radius changing from *0.42a* to *0.5a* in a two-component photonic crystal with $\varepsilon_a=1$ and $\varepsilon_b=12$.

The shift of the gaps' edges to the higher frequencies in *a)* and *b)* cases is connected with decreasing of the effective dielectric constant of the system, since the medium of DC *12* is replaced by the medium of DCs *1* and *2*, respectively (Fig.1.). This conclusion follows from the fact that the Fourier coefficients of the dielectric function are inversely proportional to the square of frequency in Maxwell equations in the reciprocal space. So, increasing of the dielectric functions can be represented as decreasing of the eigenfrequencies. To confirm this conclusion, the photonic band structures of two photonic crystals of the same symmetry but with dielectric constants differing in two times (*1* and *10*, and *2* and *20*) were calculated. All frequencies (the scale of the ordinate axis) in the latter case really became $\sqrt{2}$ smaller than in the former one, at the same time the view of the dispersion pattern was not changed. Such scaling shift can be achieved by varying either dielectric constants or filling fractions of the components, since the Fourier coefficients contain both quantities (1).

The main difference between Fig. 2a and Fig. 2b is that in the latter case the TM gap does not vanish to the close-packing condition $R+d=0.5a$. The fact is TM polarized electromagnetic wave (electrical field vector oscillates in the plane perpendicular to the cylinder's axis) "feels" whether air regions are separated by high-index dielectric or not. It is necessary to note that in a 2D photonic crystal consisting of high-index "atoms" embedded in a low-index background, the TM gaps are smaller and conditions to their appearance are stricter than in the system we studied [3,6].

The Fig. 2c describes the case of interlayer with dielectric constant $\varepsilon_i=24$. Due to the scale transformation described above the gaps' edges are shifted to the low frequencies; TE gap quickly decreases and vanishes whereas the TM gap changes slightly. Note, that all three plots start with equal frequencies of the gaps' edges at the point $d=0$.

As a matter of fact, Fig. 2b simulates the changing of the lowest band gaps' position of the macroporous silicon structure with growing silica interlayer on the surfaces of the pores. It is clear that even the interlayer of the thickness $d=0.01a$ influences the band gaps' position. We have estimated quantitatively the shift of the gaps' edges position for some "real" structures. The lattice constant $a=0.5$ μm was taken from [14]. For this case, Fig. 2b corresponds to the photonic crystal with radius $R=0.21$ μm and the interlayer thickness $d$ changing from $0$ to $40$ nm. The values of the TE and TM gaps' edges both with and without a thin interlayer of dielectric constant $\varepsilon_i=2$ in the structure with $a=0.5$ μm, $R=0.21$ μm are presented in Table 1. Relative shifts of the band gaps' edges in comparison with the case of no interlayer have magnitudes near *1.5-3%* for the interlayer thickness $d=0.01a$ and *4-7%* for $d=0.02a$. As one can see from Table 1, the absolute values of the shifts are over several tens of nanometers.

The silicon surface usually contains the native oxide layer with the thickness of the order of a few nanometers and dielectric constant that does not differ from the bulk silica [22]. The shift of the gap edges in a two-dimensional silicon photonic crystal due to the Kerr effect achieved in [14], was about *30 nm,* that is of the same order with the shifts shown in Table 1. Thus, in some cases, especially for the structures with small lattice pitch (*a<1 μm*), correction for existence of the surface specific layer should be taken into consideration.

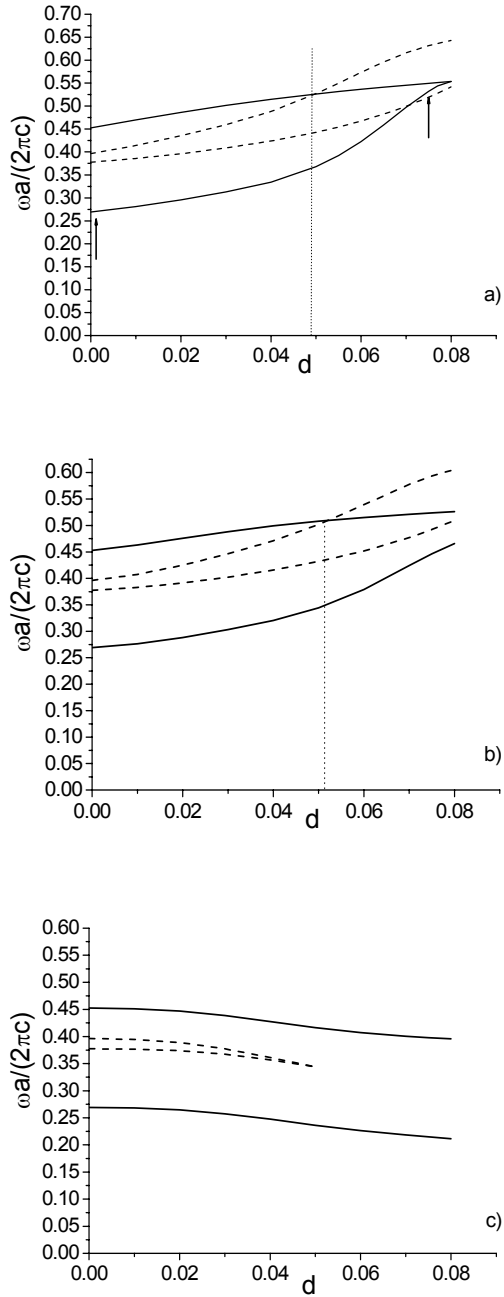

Fig. 2. TE(dashed) and TM(solid) gaps' edges position vs thickness of additional interlayer. R=0.42, $\varepsilon_a$=1, $\varepsilon_b$=12, N=729. a)$\varepsilon_i$=1, b) $\varepsilon_i$=2, c) $\varepsilon_i$=24. Vertical dotted lines show the cases of maximal absolute band gap width. The arrows correspond to the points d=0(left) and d=0.075a (right).

| Table 1. The positions of TM and TE gaps' edges for different values of interlayer thickness. The system parameters: a=0.5 μm, R=0.21 μm ε$_a$=1, ε$_b$=12, ε$_i$=2, N=1519 ||||||
|---|---|---|---|---|
|  | **TM$_{lower}$, μm** | **TM$_{upper}$, μm** | **TE$_{lower}$, μm** | TE$_{upper}$, μm |
| d=0 | 1.11 | 1.87 | 1.27 | 1.33 |
| d=5 nm | 1.09 | 1.82 | 1.24 | 1.31 |
| d=10 nm | 1.05 | 1.75 | 1.18 | 1.28 |

It is well-known [3,6] that maximal band gap width in triangular lattice 2D photonic crystals is achieved at air filling fractions higher than $f_a$=0.8. Such a big porosity may be obtained by subsequent thermal oxidation of the pore walls and etching of the oxide in hydrofluoric acid [18]. But in this case, the shortest distance between the air regions is less than *l=0.06a,* that, in turn, may influence the mechanical strength of the pattern. The maximal absolute band gap width for the oxidized photonic crystal (Fig. 2b) achieved at *R=0.42a*, *d=0.052a* is *15.6 %* of the midgap frequency, while the maximal absolute band gap width for a two-component 2D photonic crystal (Fig. 2a) achieved at *R=0.47a* is *17.3%*. Our calculations show that the absolute band gap appears even for an initial pore radius *R=0.35a* and subsequent oxidization up to the oxide thicknesses *d>0.06a* and reaches the maximum value of *12%* near *d=0.117*. Thus, it is possible to use oxidized macroporous silicon as a photonic crystal. We believe that such a photonic crystal will have some advantages over a high-porous air-silicon structure, such as higher chemical and mechanical stability. At the same time, decreasing the band gap width is not sufficient.

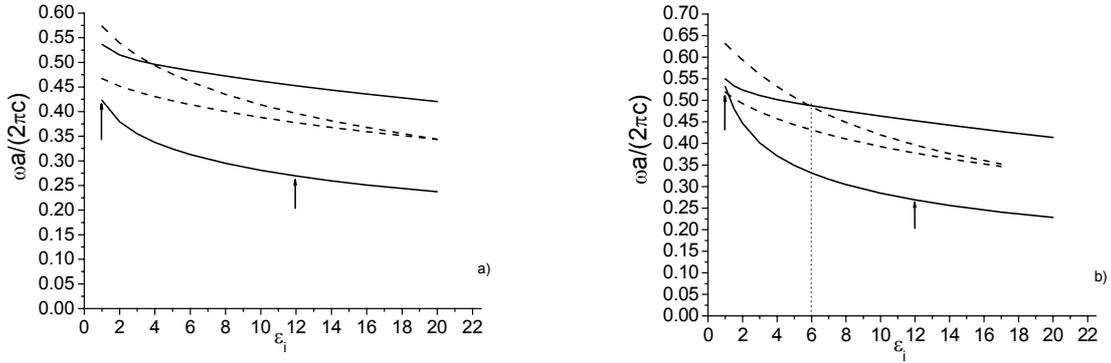

Fig. 3. TE(dashed) and TM(solid) gaps' edges position vs dielectric constant of the interlayer. R=0.42a, ε$_a$=1, ε$_b$=12, N=729. a)d=0.06a, b)d=0.075a Vertical dotted line shows the case of maximum absolute band gap width. The arrows show the points where ε$_i$=1 and ε$_i$=12

In Figs. 3a and 3b, we present the gap maps depending on the dielectric constant of the interlayer for structures with *R=0.42a, d=0.06a* and *R=0.42a, d=0.075a*, respectively. The rise of the effective DC results in a shift of the gaps to the lower frequencies. The arrows mark the points corresponding to the usual two-component photonic crystal with *ε$_i$=1* and *ε$_i$=12*. The same points can be found in Fig. 2a, too.
Thus, the left arrow in Fig. 3b shows the same point as the right arrow in Fig. 2a, and the right arrow in Fig. 3b corresponds to the same case as the left one in Fig. 2a. Generally speaking, insertion of the third component into the photonic crystal decreases the contrast of the photonic crystal and, consequently, the band gap width. But, in certain cases (like shown in Fig.3b), the presence of a third component results in widening of the absolute band gap. The maximal absolute band gap width in Fig. 3b is *11.7%* at the point *ε$_i$=6* in contrast to *17.3%* of two-component photonic crystal (Fig. 2a). In Fig. 3a the maximal absolute band gap is achieved at *ε$_i$=1*.

## 3. Conclusions

We have estimated the influence of the pore surface interlayer on the optical properties of a two-dimensional photonic crystal. It was shown that even a thin low-index interlayer ($d\sim0.01a$, $\varepsilon_i=2$) shifts the edges of the lowest gap on *1.5-3%* that may be essential for tunable photonic crystals with small lattice pitch ($a<1\ \mu m$). We propose to use the oxidized macroporous silicon with a thick oxide interlayer as a photonic crystal instead of high-porous air-silicon structures. We have shown that the insertion of a high-index interlayer insignificantly influences the TM gap edges position, while the TE gap vanishes.

It is necessary to note, that we have used the simplest approach of a real, positive, frequency-independent dielectric function. Although the main effects are included, the surface of semiconductors possesses more complicated properties, which require taking into consideration the imaginary part of the dielectric constants and their dependency on frequency.

## Acknowledgements

This work was supported by STCU grant No 2444. The authors are grateful to Prof. E. Glushko for valuable discussion.